\documentclass[12pt]{article}
\usepackage{eurosym}
\usepackage[vmargin={1in, 1in},hmargin={.9in, .9in}]{geometry}
\usepackage{latexsym}
\usepackage{amssymb}
\usepackage{amsmath}
\usepackage{amsthm}
\usepackage{mathtools}
\usepackage{amsfonts}
\usepackage{bbm}
\usepackage{graphicx}
\usepackage{enumerate}
\usepackage{color,cite}
\usepackage[breaklinks=true]{hyperref}
\usepackage{breakcites}
\usepackage{microtype}

\usepackage{setspace}
\usepackage[longnamesfirst]{natbib}
\usepackage{hypernat}

\definecolor{MyDarkBlue}{rgb}{0,0.08,0.45}
\hypersetup{linkcolor=MyDarkBlue, citecolor=MyDarkBlue, colorlinks=true}

\newtheorem{theorem}{Theorem}

\newtheorem{lemma}{Lemma}

\theoremstyle{definition}
\newtheorem{definition}{Definition}

\newtheorem{assumption}{Assumption}

\newcommand{\finexhere}{\begingroup \let\mathqed\math@qedhere
    \let\qed@elt\setQED@elt \blacktriangle@stack\relax\relax \endgroup}

\include{DDM-characterization-006.tex}

\begin{document}

\title{Testing the Drift-Diffusion Model}
\author{Drew Fudenberg \and Whitney Newey \and Philipp Strack \and Tomasz
Strzalecki}
\date{}
\maketitle

\section{Introduction}

The \emph{drift diffusion model \ }(DDM) is a model of sequential sampling
with diffusion (Brownian) signals, where the decision maker accumulates
evidence until the process hits a stopping boundary, and then stops and
chooses the alternative that corresponds to that boundary. This model has
been widely used in psychology, neuroeconomics, and neuroscience to explain
the observed patterns of choice and response times in a range of binary
choice decision problems. One class of papers study \textquotedblleft
perception tasks\textquotedblright with an objectively correct answer (e.g.
\textquotedblleft are more of the dots on the screen moving left or moving
right?\textquotedblright; here the drift of the process is related to which
choice is objectively correct \cite{Ratcliff-review,ShadlenKiani13}. The
other class of papers study \textquotedblleft consumption
tasks\textquotedblright\ such as \textquotedblleft which of these snacks
would you rather eat?\textquotedblright; here the drift is related to the
relative appeal of the alternatives %
\citep{FehrRangel,Roeetal,ClitheroRangel13,Krajbichetal10,Krajbichetal11,Krajbichetal12,Milosavljevicetal10,Krbahafe15,Reutskajaetal11}%
.

The simplest version of the DDM\ assumes that the stopping boundaries are
constant over time \cite{Wald47,Stone60,Edwards65,Ratcliff78}. More recently
a number of papers use non-constant boundaries to better fit the data, and
in particular the observed correlation between response times and choice
accuracy, i.e., that correct responses are faster than error responses \cite%
{Luce86,Milosavljevic10,Drugowitsch12,FSS2018}.

Constant stopping boundaries is the optimal solution for perception tasks
where the volatility of the signals and the flow cost of sampling are both
constant, and the prior belief is that the drift of the diffusion has only
two possible values, depending on which decision is correct. Even with
constant volatility and costs, non-constant boundaries are optimal for other
priors. \cite{FSS2018} characterize the optimal boundaries for the
consumption task: the decision maker is uncertain about the utility of each
choice, with independent normal priors on the value of each option. \cite%
{Drugowitsch12} show how to computationally derive the optimal boundaries
for the perception task: the signal coherence varies from trial to trial, so
some decision problems are harder than others.

This paper provides a statistical test for DDM's with general boundaries. We
first prove a characterization theorem: we find a condition on choice
probabilities that is satisfied if and only if the choice probabilities are
generated by some DDM. Moreover, we show that the drift and the boundary are
uniquely identified. We then use our condition to nonparametrically estimate
the drift and the boundary and construct a test statistic based on finite
samples.

Recent related work on DDM includes \cite{Drugowitsch12} who conducted a
Bayesian estimation of a collapsing boundary model and \cite{FSS2018} who
conducted a maximum likelihood estimation. \cite%
{hawkins2015revisiting} estimate collapsing boundaries in a parametric
class, allowing for a random nondecision time at the start. \cite%
{Chiongetal18} estimate a version of DDM with constant boundaries but random
starting point of the signal accumulation process; \cite{Ratcliff02}
estimates a similar model where other parameters are made random. \cite%
{Baldassi} partially characterize DDM with constant boundary.\footnote{%
They ignore the issue of correlation between response times choices by
looking only at marginal distributions, which makes their conditions
necessary but not sufficient.}

Other work on DDM-like models includes the decision field theory of \cite%
{BusemeyerTownsend,BT93,DFT} allows the signal process to be mean-reverting. 
\cite{Alosetal18} and \cite{ES17} study models where response time is a
deterministic function of the utility difference. \cite%
{HebertWoodford,Woodford14,CheMierendorff,Liangetal,LiangMu,Zhong19} study
dynamic costly optimal information acquisition.

\section{The Stochastic Choice Function}

Let $X$ be the universe of alternatives (actions) and $T=\mathbb{R}_{+}$ be
time. For every pair of objects $\{x, y\}$ the analyst observes pairwise
stochastic choices and decision times. In the limit as the sample size grows
large, the analyst will have access to the joint distribution over which
object is chosen and at which time a choice is made. We denote by $F^{xy}(t)$
the probability that the agent makes a choice by time $t$, and let $%
p^{xy}(t) $ the probability that the agent picks $x$ conditional on stopping
at time $t $. Throughout, we restrict attention to cases where $F$ has full
support and no atoms at time $0$, so that $F(0)=0$, and we assume that $F$
is strictly increasing with $\lim_{t\rightarrow \infty} F(t)=1$. These
restrictions imply the agent never stops immediately, that there is a
positive probability of stopping in every time interval, and that the agent
always eventually stops. We call $(p^{xy},F^{xy})$, the \emph{stochastic
choice function}. %
%
%  Note that every $P$
% induces a unique pair $(p,F)$ and for every pair $(p,F)$ there exists a
% unique corresponding $P$.\footnote{%
% It is given by the equation $P^{xy}(S)=\int_{0}^{\infty }\left[ p^{xy}(t)%
% \mathbf{1}_{\{(x,t)\in S\}}+(1-p^{xy}(t))\mathbf{1}_{\{(y,t)\in S\}}\right]
% dF(t)$.} Thus, we will use $P^{xy}$ and $%
% (p^{xy},F^{xy})$ interchangeably to denote the stochastic choice function associated with the pair $x,y\in
% X$.
%

An immediate restriction on the stochastic choice function is that the
choices of the agent are unaffected by which object we consider to be the
first and which object we consider to be the second. This is formally
equivalent to 
\begin{equation*}
p^{xy}(t)\equiv 1-p^{yx}(t)\text{ for all }t\text{ and }F^{xy}\equiv F^{yx}%
\text{ for all }x,y\in X.
\end{equation*}%
Without loss of generality we only consider stochastic choice functions
which satisfy this restriction. We also assume that each option is chosen
with positive probability $0<p^{xy}(t)<1$ for all $t$.

Given $(p^{xy},F^{xy})$ we define the choice imbalance at each time $t$ to
be 
\begin{equation*}
I^{xy}(t):=p^{xy}(t)\log \left( \frac{p^{xy}(t)}{1-p^{xy}(t)}\right)
+(1-p^{xy}(t))\log \left( \frac{1-p^{xy}(t)}{p^{xy}(t)}\right) \,.
\end{equation*}%
This is the Kullback-Leibler divergence (or relative entropy) between the
Binomial distribution of the agent's time $t$ choice $P(t)=(p(t),1-p(t))$
and the permuted choice distribution $Q(t)=(1-p(t),p(t))$. As the
Kullback-Leibler divergence is a statistical measure of the similarity
between distributions $I(t)$ captures the imbalance of the agent's choice at
time $t$. Note that $I=0$ means that both choices are equally likely; $%
I=\infty $ when $p$ equals $0$ or $1$, and that $I$ is symmetric about~0.5.
We define $\bar{I}^{xy}$ to be the average choice imbalance, 
\begin{equation*}
\bar{I}^{xy}:=\int_{0}^{\infty }I^{xy}(t)\,dF^{xy}(t)\,,
\end{equation*}%
and we define $\bar{T}^{xy}$ to be the average decision time, 
\begin{equation*}
\bar{T}^{xy}:=\int_{0}^{\infty }t\,dF^{xy}(t)\,,
\end{equation*}%
and define $\bar{p}^{xy}$ to be the average choice probability$,$%
\begin{equation*}
\bar{p}^{xy}:=\int_{0}^{\infty }p^{xy}(t)\,dF^{xy}(t)\,,
\end{equation*}%
and assume that all of these integrals exist. Finally, we relabel objects as
needed so that the first object is chosen weakly more often, i.e. $\bar{p}%
^{xy}\geq 0.5$ for all $x,y$.

\section{DDM representation}

The drift diffusion model (DDM) is commonly used to explain the stochastic
choice data in neuroscience and psychology. The two main ingredients of a
DDM are the stimulus process $Z_{t}$ and a time-dependent stopping boundary $%
b(t)$. In the DDM representation, the stimulus process $Z_{t}$ is a Brownian
motion with drift $\delta $ and volatility $\alpha $: 
\begin{equation}
Z_{t}=\delta \,t+\alpha \,B_{t},  \label{eq:Z}
\end{equation}%
where $B_{t}$ is a standard Brownian motion, so in particular $Z_{0}=0$.
Define the hitting time $\tau $ 
\begin{equation}
\tau =\inf \{t\geq 0:|Z_{t}|\geq b(t)\},  \label{eq:tau}
\end{equation}%
i.e., the first time the absolute value of the process $Z_{t}$ hits the
boundary $b$. Let $F^{\ast }(t;\delta ,b,\alpha ):=\mathbb{P}\left[ \tau
\leq t\right]$ be the distribution of the stopping time $\tau $. Likewise,
let $p^{\ast }(t;\delta ,b,\alpha )$ be the conditional choice probability
induced by \eqref{eq:Z} and \eqref{eq:tau} and a decision rule that chooses $%
x$ if $Z_{\tau }=b(\tau )$ and $y$ if $Z_{\tau }=-b(\tau )$.

Our goal in this paper is to determine which data is consistent with a DDM
representation, and when it is, when the representation is unique. When the
drift $\delta=0$, each alternative will be chosen half of the time
regardless of the shape of the boundary, so we will exclude this case going
forward.

The original formulation of the DDM was for \textquotedblleft perception
tasks\textquotedblright\ where the drift $\delta $ is either $+1$ or $-1$
depending on which decision is correct; more generally there can be a
distinct drift $\delta ^{xy}$ for each pair $x,y$. In consumption-choice
problems (otherwise known as value-based problems, see, e.g., \cite%
{Milosavljevic10}) it is natural to assume that the net drift $\delta ^{xy}$
is the difference between two signals, an $x$-signal with drift $u(x)$ equal
to the utility of $x$ and a $y$-signal with drift $u(y)$ equal to the
utility of $y$, so that $\delta ^{xy}=u(x)-u(y).$ This imposes some
consistency conditions that we discuss below.

\begin{definition}[DDM Representation]
Stochastic choice data $(p^{xy},F^{xy})_{x,y\in X}$ has a DDM representation
if there exists a utility function $u:X\rightarrow \mathbb{R}$, a volatility
parameter $\alpha >0$ as well as a boundary $b:\mathbb{R}_{+}\rightarrow 
\mathbb{R}_{+}$ such that for all $x,y\in X$ and $t\in \mathbb{R}$ 
\begin{align*}
p^{xy}(t)&=p^{\ast }\Big(t,u(x)-u(y),b,\alpha \Big) \\
\text{ and }F^{xy}(t)&=F^{\ast }\Big(t,u(x)-u(y),b,\alpha \Big)\,.
\end{align*}
\end{definition}

Note that this definition requires that the data from all of the menus $%
\{x,y\}$ is generated with the \emph{same} boundary function $b$. This
corresponds to cases where the agent treats each decision problem as a
random draw from a fixed environment.\footnote{%
In an optimal stopping model, the shape of the boundary is determined by the
agent's prior over these draws.} We are interested in characterizing which
stochastic choice functions admits a DDM representation. The following
result follows immediately from rescaling $\delta $ and $b$.

\begin{lemma}
\label{lem:alpha} If a stochastic choice function exhibits a DDM
representation for some $\alpha $, then it also exhibits a DDM
representation for $\alpha =1$.
\end{lemma}

We will thus without loss of generality only consider the DDM model where we
normalized $\alpha =1$. We write $p^{\ast }(t,\delta ,b)$ and $F^{\ast
}(t,\delta ,b)$ as short-hands for $p^{\ast }(t,\delta ,b,1)$ and $F^{\ast
}(t,\delta ,b,1)$.

\section{Characterization}

\label{sec:theory}

Given a stochastic choice function $(p^{xy},F^{xy})$, define the \emph{%
revealed drift} 
\begin{equation}
\widetilde{\delta }^{xy}:=\sqrt{\frac{\bar{I}^{xy}}{2\bar{T}^{xy}}}.
\label{eq:delta}
\end{equation}%
When the revealed drift is is non zero, we define the \emph{revealed boundary%
} as 
\begin{equation}
\widetilde{b}^{xy}(t):=\frac{\ln p^{xy}(t)-\ln (1-p^{xy}(t))}{2\widetilde{%
\delta }^{xy}}.  \label{eq:b}
\end{equation}
The revealed drift is high for a pair $x,y$ whenever the agent either makes
very imbalanced choices or decides quickly, and low for choices that are
slow and close to 50-50. Over time the boundary at time $t$ follows the
log-odds ratio of the agent's choice at time $t$ which is zero whenever the
agent's choice is balanced and and increases in the imbalance of the agent's
choice. The revealed boundary is smaller for pairs with a larger revealed
drift. In the knife-edge case when the revealed drift is 0 the revealed
boundary is not defined and our results do not apply.

Theorem \ref{thm:1} below says that if the true data generating process is a
DDM, then the revealed drift and boundary will exactly match the true
parameters. Moreover, Theorem \ref{thm:1} allows us to test whether the true
data generating process is indeed a DDM.

\subsection{Characterization for a fixed pair}

Our first result characterizes the DDM for a fixed pair $x, y\in X$.

\begin{theorem}
\label{thm:1} For a fixed pair $x,y$ with $\tilde{\delta}%
^{xy}\neq 0$ the stochastic choice function $(p^{xy},F^{xy})$
admits a DDM representation if and only if for all $t\geq 0$ 
\begin{equation*}
F^{xy}(t)=F^{\ast }(t;\tilde{\delta}^{xy},\tilde{b}^{xy})
\end{equation*}%
If such a representation exists it is unique (up to the choice of $\alpha $)
and given by $\tilde{\delta}^{xy},\tilde{b}^{xy}$.
\end{theorem}

Thus, the stochastic choice function $(p^{xy},F^{xy})$ is consistent with
DDM whenever the observed distribution of stopping times $F^{xy}$ equals to
the distribution of hitting times generated by the revealed drift $\tilde{%
\delta}^{xy}$ and revealed boundary $\tilde{b}^{xy}$. Theorem \ref{thm:1}
shows that the revealed drift and boundary are the unique candidate for a
DDM representation. It thus allows us to identify the parameters of the DDM
model directly from choice data. This permits the model to be calibrated to
the data without computing the likelihood function, which requires
computationally costly Monte-Carlo simulations. More substantially, as
Theorem \ref{thm:1} connects the primitives of the model directly to data it
allows us to better understand their economic meaning. The drift in the DDM
model is a measure of how imbalanced and quick the agent's choices are and
the shape of the boundary follows the imbalance of the agent's choices over
time. We hope that this interpretation makes the empirical content of the
parameters of DDM model more transparent and the model thus more useful.

Note that this theorem shows that the distribution of stopping times
contains additional information that is not captured by the mean. For
example, choice data where $p^{xy}(t)$ and $\bar{T}^{xy}$ are any 2 given
constants is only consistent with one possible distribution of stopping
times $F^{xy}$\ \ However a test based only on the mean choice probability
and mean stopping time will accept any model that matches those two numbers,
and in particular regardless of $F^{xy}$ the data is consistent with a
constant stopping boundary. (See \cite{Baldassi}).

\subsection{Characterization for menus of pairs}

Our next result extends the characterization to all pairs $x, y\in X$.

\begin{theorem}
\label{thm:2} The stochastic choice function $(\{p^{xy}\},\{F^{xy}\})_{x,y%
\in X})$ has a DDM representation iff

\begin{enumerate}
\item[(i)] $F^{xy}(t)=F^{*}(t; \tilde \delta^{xy}, \tilde b^{xy})$ for all $%
t\geq0$,

\item[(ii)] $\widetilde{b}^{xy}(t)=\widetilde{b}^{xz}(t)$ for all $x,y,z\in X
$ and all $t\geq 0$.

\item[(iii)] $\widetilde{\delta }^{xy}+\widetilde{\delta }^{yz}=\widetilde{%
\delta }^{xyz}$ for all $x,y,z\in X$,
\end{enumerate}
\end{theorem}

Thus, in addition to satisfying the condition from Theorem 1 pairwise, we
have two additional consistency conditions imposed across pairs. Condition
(ii) follows from our assumption that the agent uses the same stopping
boundary in every menu. Condition (iii) comes from the assumption that the
drift in a given menu depends on the difference of utilities, that is $%
\delta ^{xy}=u(x)-u(y)$.\footnote{%
The proof of the theorem follows from Theorem \ref{thm:1} and the Sincov
functional equation, see, e.g., \cite{Aczel66}.}

\section{An Econometric Test for a Fixed Pair of Alternatives}

\label{sec:metrics}

The idea for the test is based on Theorem \ref{thm:1}, which requires that
the observed distribution of stopping times matches the distribution induced
by the revealed boundary $\tilde{b}$ and drift $\tilde{d}$. We first
describe a nonparametric estimator of $\tilde{b}$ and $\tilde{\delta}$ based
on a finite data set. Next, we show how to test the distribution matching
condition. This test could be extended to multiple-alternatives settings
along the lines of Theorem \ref{thm:2}, but we do not do so here.

\subsection{Estimation of drift and boundary}

Suppose that we have a fixed pair $x,y\in X$. Define 
\begin{equation*}
\gamma _{\tau }:=\left\{ 
\begin{array}{ll}
1, & \text{when choice $x$ is made,} \\ 
0, & \text{when choice $y$ is made.}%
\end{array}%
\right.
\end{equation*}%
Each data point consists of the time $\tau _{i}$ at which the choice is made
and the choice $\gamma _{i}$ made at time $\tau _{i}.$

\begin{assumption}
The data $(\tau _{1},\gamma _{1}),\ldots ,(\tau _{n},\gamma _{n})$ are i.i.d.
\end{assumption}

The unknown features of the DDM model are the drift $\delta $ and the
boundary $b(t).$\textbf{\ }We use estimators based on equations %
\eqref{eq:delta} and \eqref{eq:b} that identify the revealed drift and
boundary. Both of them depend on the choice probability, so we first give an
estimator of that. Here $p^{xy}\left( t\right) :=\Pr (\gamma _{i}=1|\tau
_{i}=t)$ is the probability of choice $x$ conditional on the choice being
made at $t$.

The nonparametric estimator we construct is a spline regression: that is, a
least squares regression of $\gamma _{i}$ on approximating functions of $%
\tau _{i}.$ For simplicity, we use a linear probability estimator of $%
p^{xy}(t)$.\footnote{%
We reserve consideration of other estimators of the choice probability to
future work, including logit or probit with a series approximation inside
the logit or probit CDF.}

We first transform $\tau _{i}$ to the unit interval.\footnote{%
In DDM models where $b(t)$ does not reach zero, there is no uniform bound on
realized decision times $\tau _{i}$. Because $\tau _{i}$ is the conditioning
variable (i.e. regressor) in the choice probability, it is important to
allow for an unbounded regressor.} For this purpose let $G(t)$ be a CDF of a
positive random variable with PDF that is positive on $(0,\infty ).$ Consider%
\begin{equation*}
G_{i}=G\left( \tau _{i}\right) .
\end{equation*}%
Because $G_{i}$ lies in the unit interval we can use standard series
estimation to estimate $p^{xy}(t)$. We consider regression spline estimation
of $p^{xy}(t)$. For this purpose let 
\begin{equation*}
q^{K}\left( G\right) =\left( q_{1K}\left( G\right) ,\ldots ,q_{KK}\left(
G\right) \right) ^{\prime }
\end{equation*}%
be a $B$-spline vector, say for evenly spaced knots on $(0,1)$. Let $\hat{%
\beta}$ be OLS coefficients from regressing $\gamma _{i}$ on $%
q_{i}^{K}=q^{K}\left( G_{i}\right) $. The choice probability estimator we
consider is%
\begin{equation*}
\hat{p}\left( t\right) :=q^{K}\left( G(t)\right) ^{\prime }\hat{\beta},\text{
}\hat{\beta}:=\left[ \sum_{i=1}^{n}q_{i}^{K}q_{i}^{K}{}^{\prime }\right]
^{-1}\sum_{i=1}^{n}q_{i}^{K}\gamma _{i}.
\end{equation*}%
We give conditions for this estimator to be consistent and have other
important large sample properties in Assumptions 2 and 3 to follow.

We can estimate the drift $\delta $ by plugging in $\hat{p}(t)$ for $%
p^{xy}(t)$ in formula \eqref{eq:delta} and replacing expectations with
sample averages. Let 
\begin{align*}
\hat{I}(t)& :=\hat{p}\left( t\right) \ln \left[ \frac{\hat{p}\left( t\right) 
}{1-\hat{p}\left( t\right) }\right] +\left[ 1-\hat{p}\left( t\right) \right]
\ln \left[ \frac{1-\hat{p}\left( t\right) }{\hat{p}\left( t\right) }\right] ,
\\
\bar{I}& :=\frac{1}{n}\sum_{i=1}^{n}\hat{I}\left( \tau _{i}\right) ,\text{ }%
\bar{\tau}:=\frac{1}{n}\sum_{i=1}^{n}\tau _{i}.
\end{align*}%
The estimator of $\delta $ is then%
\begin{equation*}
\hat{\delta}:=\sqrt{\frac{\bar{I}}{\bar{\tau}}}.
\end{equation*}%
The estimator of the boundary $b\left( t\right) $ is obtained by plugging in 
$\hat{\delta}$ and $\hat{p}(t)$ in the expression of equation \eqref{eq:b},
giving

\begin{equation*}
\hat{b}(t):=\frac{1}{\hat{\delta}}\ln \left[ \frac{\hat{p}\left( t\right) }{%
1-\hat{p}\left( t\right) }\right] .
\end{equation*}

\subsection{Testing}

We now have to test whether the observed distribution of stopping times
matches the one induced by the revealed drift and boundary. We do this by
comparing sample moments of functions of the decision time with estimators
of the moments that predicted by the model. To describe such a test let $%
m_{J}(\tau )=(m_{1J}(\tau ),...,m_{JJ}(\tau ))^{\prime }$ be a vector of
functions of $\tau $. Examples include indicator functions for intervals and
B-splines in $G(\tau ).$ The sample average vector will be $\bar{m}%
=\sum_{i=1}^{n}m_{J}(\tau _{i})/n$.\footnote{%
The Kolmogorov--Smirnoff test uses indicator functions but instead of the
the average of $m$ it takes the supremum. The Cramer--von Mises test takes
the sum of squares. We look at the average of $m$ because the target cdf we
are comparing with is not fixed, but involves estimates of the boundary and
drift, see \cite{Newey94}.} We use simulation to obtain model prediction. To
describe the simulated predictions, let $\{B_{t}^{1},...,B_{t}^{S}\}$ be $S$
independent copies of Browning motion and 
\begin{equation*}
\hat{\tau}_{s}=\inf \{t\geq 0:\big\vert\hat{\delta}t+B_{t}^{s}\big\vert\geq 
\hat{b}(t)\}.
\end{equation*}%
A moment vector predicted by the model would be $\hat{m}_{S}=%
\sum_{s=1}^{S}m_{J}(\hat{\tau}_{s})/S.$ A test of the model can be based on
comparing $\bar{m}$ and $\hat{m}.$ Let $\hat{V}$ be a consistent estimator
of the asymptotic variance of $\sqrt{n}(\bar{m}-\hat{m}_{S})$ when the model
is correctly specified, as we will describe below. A test statistic can be
formed as%
\begin{equation*}
\hat{A}:=n(\bar{m}-\hat{m}_{S})^{\prime }\hat{V}^{-1}(\bar{m}-\hat{m}_{S}).
\end{equation*}%
The model would be rejected if $\hat{A}$ exceeds the critical value of a $%
\chi ^{2}(J)$ distribution. If $J$ is allowed to grow with $n$ and the $%
m_{J}(\tau )$ is allowed to grow in dimension and richness as $n$ grows then
this approach will test all the restrictions implied by DDM as $n$ grows. In
Appendix A we describe the construction of $\hat{V}.$

In formulating conditions for the asymptotic distribution of this test we
will let $m_{jJ}(\tau ),$ $(j=1,...,J)$ be indicator functions for disjoint
intervals. Let $\tau _{jJ}=G^{-1}(j/(J+1)),$ $(j=0,...,J),$ $\tau
_{J+1,J}=\infty .$ Consider 
\begin{equation*}
m_{jJ}(t)=\sqrt{J+1}\cdot \mathbbm{1}(\tau _{j,J}\leq t<\tau _{j+1,J}),\text{
}(j=1,...,J).
\end{equation*}%
The test based on these functions will be based on comparing empirical
probabilities of intervals with those predicted by the model. The
normalization of multiplying by $\sqrt{J+1}$ is convenient in making the
second moment of these functions of the same magnitude for different values
of $J$. Note that we have left out the indicator for the interval $%
(0,1/(J+1)).$ We have done this to account for the fact that the estimator
the drift parameter uses some information about $\tau _{i}$, so that we are
not able to test all of the implications of the DDM for the distribution of $%
\tau _{i}.$ As usual we can only test overidentifying restrictions.

We derive results under the following conditions:

\begin{assumption}
\label{asst:2} The pdf of $G(\tau _{i})$ is bounded and bounded away from
zero.
\end{assumption}

This assumption is equivalent to the ratio of the pdf of $\tau _{i}$ to $%
dG(t)/dt$ being bounded and bounded away from zero. It is straightforward to
weaken this condition to allow it to only hold on compact, connected
interval that is a subset of $(0,1),$ if we assume the $b(t)$ is constant on
known intervals near $0$ and where $\tau $ is large. 

We also make a smoothness assumption on the boundary function.

\begin{assumption}
\label{asst:3} $b(G^{-1}(g))$ is bounded and $s\geq 1$ times differentiable
with bounded derivatives on $g\in \lbrack 0,1]$ and the $q_{kK}(G),$ $%
k=1,...,K$ are b-splines of order $s-1.$
\end{assumption}

This condition requires that the derivatives of $b(t)$ go to zero in the
tails of the distribution of $\tau _{i}$ as fast as the pdf of $G(t)$ does.
We also require that the drift parameter be nonzero.

\begin{assumption}
\label{asst:4} $\delta \neq 0$.
\end{assumption}

We need to add other conditions about the smoothness of CDF of $\tau _{i}$
as a function of the drift $\delta $ and the boundary and about rates of
growth of $J$ and $K$. The involve much notation, so we state them in
Assumption \ref{asst:5} in Appendix \ref{app:smoothness}.

We can now state the following result on the limiting distribution of $\hat{A%
}.$

\begin{theorem}
Suppose that Assumptions \ref{asst:2}, \ref{asst:3}, \ref{asst:4} and
Assumption \ref{asst:5} in Appendix \ref{app:smoothness} are satsified. Then
for the $1-\alpha $ quantile $c\left( \alpha ,J\right) $ of a chi-square
distribution with $J$ degrees of freedom%
\begin{equation*}
\Pr \left( \hat{A}\geq c\left( \alpha ,J\right) \right) \longrightarrow
\alpha .
\end{equation*}
\end{theorem}

\appendix

\section{Proofs from Section \protect\ref{sec:theory}}

\label{app:theory}

\subsection{Proof of Lemma \protect\ref{lem:alpha}}

Dividing (\ref{eq:Z}) by $\alpha$ and observing that $\inf \{t\geq
0:|Z_{t}|\geq b(t)\} = \inf \{t\geq 0:|\frac{Z_{t}}{\alpha}|\geq \frac{b(t)}{%
\alpha}\}$ yields that $p^{*}\Big(\delta(x,y),b,\alpha\Big) =p^{*}\Big(\frac{%
1}{\alpha}\delta(x,y),\frac{b}{\alpha},\alpha\Big)$ and thus the result.\qed%
\newline

\subsection{Proof of Theorem \protect\ref{thm:1}}

(1) We first show that these conditions are necessary for $(p,F)$ to admit a
DDM representation for a given pair $\{x,y\}.$

By equation (4) in \cite{FSS2018} we have $\frac{p^{xy}(t)}{1-p^{xy}(t)}%
=\exp \left(2 \delta (x,y)\,b(t)\right) $. Thus, we have that 
\begin{equation}
b(t)=\frac{1}{2\delta (x,y)}\log \left( \frac{p^{xy}(t)}{1-p^{xy}(t)}\right)
\,.  \label{eq:conditional-choice-prob}
\end{equation}%
This proves \eqref{eq:b}. By the definition of $\tau $ in equation (\ref{eq:tau}) we have $Z_{\tau }=%
\text{sgn}(Z_{\tau })b(\tau )$. By (\ref{eq:Z}), $Z_{\tau }=\delta (x,y)\tau
+\,B_{\tau }$. Combining these two equations and taking
expectations, it follows from Doob's optional sampling theorem that 
\begin{equation}
\delta (x,y)\,\mathbb{E}^{xy}\left[ \tau \right] =\mathbb{E}^{xy}\left[ 
\text{sgn}(Z_{\tau })b(\tau )\right]  \label{eq:avg-stopping-time}
\end{equation}%
Plugging (\ref{eq:conditional-choice-prob}) into (\ref{eq:avg-stopping-time}%
) yields 
\begin{equation*}
\delta (x,y)\,\mathbb{E}^{xy}\left[ \tau \right] =\mathbb{E}^{xy}\left[ 
\text{sgn}(Z_{\tau })\frac{1}{2\delta (x,y)}\log \left( \frac{p^{x}(\tau )}{%
1-p^{xy}(\tau )}\right) \right]
\end{equation*}%
Dividing by $\mathbb{E}^{xy}\left[ \tau \right] $ and multiplying by $2\delta
(x,y)$ yields 
\begin{align*}
2\delta (x,y)^{2}& =\frac{\mathbb{E}^{xy}\left[ \text{sgn}(Z_{\tau })\log
\left( \frac{p^{x}(\tau )}{1-p^{xy}(\tau )}\right) \right] }{\mathbb{E}^{xy}%
\left[ \tau \right] }=\frac{\mathbb{E}^{xy}\left[ [\mathbf{1}_{Z_{\tau }>0}-%
\mathbf{1}_{Z_{\tau }<0}]\log \left( \frac{p^{x}(\tau )}{1-p^{xy}(\tau )}%
\right) \right] }{\mathbb{E}^{xy}\left[ \tau \right] } \\
& =\frac{\mathbb{E}^{xy}\left[ \int_{0}^{\infty }\mathbf{1}_{\tau=t}[\mathbf{%
1}_{Z_{\tau }>0}-\mathbf{1}_{Z_{\tau }<0}]\log \left( \frac{p^{xy}(t)}{%
1-p^{xy}(t)}\right) \mathrm{d}t\right]}{\int_{0}^{\infty }t\,\,\mathrm{d}%
F^{xy}(t)} \\
& =\frac{\mathbb{E}^{xy}\left[ \int_{0}^{\infty }\mathbf{1}_{\tau=t} \mathbb{%
E}^{xy} \left[[\mathbf{1}_{Z_{\tau }>0}-\mathbf{1}_{Z_{\tau }<0}]\log \left( 
\frac{p^{xy}(t)}{1-p^{xy}(t)}\right) \mid \tau = t\right]\mathrm{d}t\right]}{%
\int_{0}^{\infty }t\,\,\mathrm{d}F^{xy}(t)} \\
& =\frac{\mathbb{E}^{xy}\left[ \int_{0}^{\infty }\mathbf{1}_{\tau=t} [%
\mathbb{E}^{xy} \left[\mathbf{1}_{Z_{\tau }>0} \mid \tau = t\right]-\mathbb{E%
}^{xy}\left[\mathbf{1}_{Z_{\tau }<0}\mid\tau=t\right]]\log \left( \frac{%
p^{xy}(t)}{1-p^{xy}(t)}\right)\mathrm{d}t\right]}{\int_{0}^{\infty }t\,\,%
\mathrm{d}F^{xy}(t)} \\
& =\frac{\mathbb{E}^{xy}\left[ \int_{0}^{\infty }\mathbf{1}_{\tau=t}
[p^{xy}(t)-(1-p^{xy}(t))]\log \left( \frac{p^{xy}(t)}{1-p^{xy}(t)}\right)%
\mathrm{d}t\right]}{\int_{0}^{\infty }t\,\,\mathrm{d}F^{xy}(t)} \\
& =\frac{\int_{0}^{\infty }[p^{xy}(t)-(1-p^{xy})(t)]\log \left( \frac{%
p^{xy}(t)}{1-p^{xy}(t)}\right) \mathrm{d}F^{xy}(t)}{\int_{0}^{\infty }t\,\,%
\mathrm{d}F^{xy}(t)} \\
& =\frac{\int_{0}^{\infty }[2\,p^{xy}(t)-1]\log \left( \frac{p^{xy}(t)}{%
1-p^{xy}(t)}\right) \mathrm{d}F^{xy}(t)}{\int_{0}^{\infty }t\,\,\mathrm{d}%
F^{xy}(t)}\,.
\end{align*}

This proves \eqref{eq:delta}. Finally, we know that $\delta >0$ if and only if the probability with which
the first object is chosen $\mathbb{P}^{xy}[Z_{\tau }>0]=\int_{0}^{\infty
}p^{xy}(t)\mathrm{d}F^{xy}(t)$ is greater $\frac{1}{2}$ which yields the
result.

To show sufficiency, consider the DDM model with parameters $(\tilde{\delta}%
^{xy},\tilde b^{xy})$ given by (\ref{eq:delta}--\ref{eq:b}). It follows that $F^{xy}$ equals the
distribution over stopping times in the DDM model with boundary $\tilde{b}%
^{xy}$ and drift $\delta ^{xy}$. Finally, we will show that this DDM model
also generates the correct conditional stopping probabilities $p^{xy}$. By
equation (4) in \cite{FSS2018}, the conditional probability of stopping in
the DDM model $\tilde{p}^{xy}$ satisfies 
\begin{equation*}
\frac{\tilde{p}^{xy}(t)}{1-\tilde{p}^{xy}(t)}=\exp \left( 2\tilde{\delta}%
(x,y)\,\tilde{b}^{xy}(t)\right) =\frac{{p}^{xy}(t)}{1-{p}^{xy}(t)}\,,
\end{equation*}%
which completes the proof as we have argued that each stochastic choice
function is uniquely identified by the associated pair $(p,F)$.\qed

\section{Construction of $\hat V$}

\label{app:Vhat}

To construct $\hat{V}$ we use the fact that there are three asymptotically
independent sources of variation in $\bar{m}-\hat{m}$. These sources are the
variation in $\tau _{i},$ the variation in $\hat{\beta}$, and the variation
from simulation. The variation in $\tau _{i}$ affects both $\bar{m}$ and $%
\hat{\delta}$ and the variation in $\hat{\delta}$ has an effect through $%
\hat{m}.$ Generally $\hat{m}$ will not be differentiable in $\hat{\delta}$
so we use a difference quotient to estimate the derivative of $\hat{m}$ with
respect to $\delta .$ To describe how this source of variation can be
estimated let%
\begin{equation*}
\tau _{s}(\delta ,\beta )=\inf \{t\geq 0:\left\vert \delta
t+B_{t}^{s}\right\vert \geq \frac{1}{\delta }\ln \left[ \frac{%
q^{K}(G(t))^{\prime }\beta }{1-q^{K}(G(t))^{\prime }\beta }\right] \},\text{ 
}\hat{m}(\delta ,\beta )=\frac{1}{S}\sum_{s=1}^{S}m_{J}(\tau _{s}(\delta
,\beta )).
\end{equation*}%
denote one simulation $\tau _{s}(\delta ,\beta )$ of $\tau _{s}$ when $%
\delta $ is the true drift and $q_{K}(G(t))^{\prime }\beta $ the true $%
p(t)=p^{xy}(t)$ and $\hat{m}(\delta ,\beta )$ denote the average over $S$
simulations. Let

\begin{equation*}
\hat{M}_{\delta }=\frac{\hat{m}(\hat{\delta}+\Delta ,\hat{\beta})-\hat{m}(%
\hat{\delta}-\Delta ,\hat{\beta})}{2\Delta }
\end{equation*}%
be the difference quotient that serves as an estimator of the derivative of
the the expectation of the model moments with respect to the drift. Then%
\begin{equation*}
\hat{\psi}_{i1}=m_{J}(\tau _{i})-\bar{m}-\hat{M}_{\delta }\frac{1}{2\hat{%
\delta}\bar{\tau}}[\hat{I}(\tau _{i})-\bar{I}-\hat{\delta}^{2}\{\tau _{i}-%
\bar{\tau}\}]
\end{equation*}%
will estimate the influence of $\tau _{i}$ on the difference of moments
coming from the effect of $\tau _{i}$ on the sample moments as well as on $%
\hat{\delta}.$ An estimator of the variance of the moment differences due to
variation in $\tau _{i}$ is then%
\begin{equation*}
\hat{V}_{1}=\frac{1}{n}\sum_{i=1}^{n}\hat{\psi}_{i1}\hat{\psi}_{i1}^{\prime
}.
\end{equation*}

To estimate the component of the variance due to $\hat{\beta}$ we use%
\begin{equation*}
\hat{M}_{k}=\frac{\hat{m}(\hat{\delta},\hat{\beta}+e_{k}\Delta )-\hat{m}(%
\hat{\delta},\hat{\beta}-e_{k}\Delta )}{2\Delta },\text{ \ }\hat{M}_{\beta
}=[\hat{M}_{1},...,\hat{M}_{K}].
\end{equation*}%
to estimate the derivative of $E[m_{J}(\tau _{s}(\delta ,\beta ))]$ with
respect to $\beta $ at $\hat{\delta}$ and $\hat{\beta},$ where $e_{k}$ is
the $k^{th}$ unit vector. Let $\hat{p}_{i}=\hat{p}(\tau _{i})$ and $%
d(p)=d\ln [p/(1-p)]/dp=p^{-1}(1-p)^{-1}$. Accounting also for the effect of $%
\beta $ on $\hat{\delta}$, an estimator of the Jacobian of $E[m_{J}(\tau
_{s}(\delta ,\beta ))]$ with respect to $\beta $ is

\begin{equation*}
\hat{D}_{\beta }=\hat{M}_{\delta }\frac{1}{2\hat{\delta}\bar{\tau}n}%
\sum_{i=1}^{n}d(\hat{p}_{i})q_{i}^{K}{}^{\prime }+\hat{M}_{\beta }.
\end{equation*}%
The variation in $\bar{m}-\hat{m}$ due to $\hat{\beta}$ can then be
estimated by%
\begin{equation*}
\hat{V}_{2}=\hat{D}_{\beta }\hat{\Sigma}^{-1}\left[ \frac{1}{n}%
\sum_{i=1}^{n}q_{i}^{K}q_{i}^{K\prime }(\gamma _{i}-\hat{p}_{i})^{2}\right] 
\hat{\Sigma}^{-1}\hat{D}_{\beta }^{\prime },\text{ }\hat{\Sigma}=\frac{1}{n}%
\sum_{i=1}^{n}q_{i}^{K}q_{i}^{K\prime }.
\end{equation*}%
This is a delta method estimator of the asymptotic variance of $E[m_{J}(\tau
_{s}(\delta ,\beta ))]$ due to the $\hat{\beta}$ in the nonparametric
estimator $\hat{p}(t)$. As in \cite{Newey94}, it is formed by treating $\hat{%
m}$ as depending on the vector of parameters $\hat{\beta}$ and applying the
delta method as if $K$ were fixed and not growing with the sample size.

The variation due to simulation is easy to estimate as $\hat{V}%
_{3}=(n/S^{2})\sum_{s=1}^{S}\left[ m_{J}(\hat{\tau}_{s})-\hat{m}\right] %
\left[ m_{J}(\hat{\tau}_{s})-\hat{m}\right] ^{\prime }.$ In the theory we
assume that the number of simulations is large enough so that we can replace
this $\hat{V}_{3}$ by zero without affecting the results. Computing $\hat{V}%
_{3}$ in practice may still be a good idea check whether the number of
simulations is large enough to make $\hat{V}_{3}$ negligible.

The estimators of the variance from independent sources of variation can
then be combined into an asymptotic variance estimator for $\sqrt{n}[\bar{m}-%
\hat{m}_{S}]$ as%
\begin{equation*}
\hat{V}=\hat{V}_{1}+\hat{V}_{2}+\hat{V}_{3}.
\end{equation*}%
We give conditions in Theorem 3 sufficient for the chi-squared approximation
to the distribution of $\hat{A}$ to be correct for $n,$ $J$, and $S$ growing
and $\Delta $ shrinking in specific ways.

\section{Smoothness Conditions for the CDF of $\protect\tau _{i}. $}

\label{app:smoothness}

To obtain the limiting distribution of the test statistic we make use of
smoothness conditions for the CDF of $\tau _{i}$ as $F(t|\delta ,b)$ as a
function of the drift $\delta $ and boundary $b(\cdot )$. The three key
primitive regularity conditions that will be useful involve a Frechet
derivative $D(\tilde{\delta}-\delta ,\tilde{b}-b;\delta ,b,t)$ of $%
F(t|\delta ,b)$ with respect to $\delta $ and $b$. We collect these
conditions in the following assumption. Let $\varepsilon _{pn}=\sqrt{%
n^{-1}K\ln (K)/n}+K^{-s}.$

\begin{assumption}
\label{asst:5}

For $\left\vert \tilde{b}\right\vert =\sup_{t}\left\vert \tilde{b}%
(t)\right\vert $ there is $C>0$ not depending on $\delta $, $b$, $t$ such
that

a) 
\begin{equation*}
|F(t|\tilde{\delta},\tilde{b})-F(t|\delta ,b)+D(\tilde{\delta}-\delta ,%
\tilde{b}-b;\delta ,b,t)|\leq C(|\tilde{\delta}-\delta |^{2}+|\tilde{b}%
-b|^{2});
\end{equation*}

b) for each $t$ there is a constant $D_{0t}^{\delta }$ and function $\alpha
_{0t}(t)$ such that $\left\vert \alpha _{0t}(\tau _{i})\right\vert \leq C$, $%
\left\vert D_{0t}^{\delta }\right\vert \leq C$, $\left\vert d^{s}\alpha
_{0t}(t)/dt^{s}\right\vert \leq C$ for $s$ equal to the order of the spline
plus $1$, and%
\begin{equation*}
D(\tilde{\delta}-\delta ,\tilde{b}-b;\delta ,b,t)=D_{0t}^{\delta }(\tilde{%
\delta}-\delta )+E[\alpha _{0t}(\tau _{i})\{\tilde{b}(\tau _{i})-b(\tau
_{i})\}];
\end{equation*}

c) 
\begin{equation*}
|D(\delta ,b;\tilde{\delta},\tilde{b},t)-D(\delta ,b;\delta
_{0},b_{0},t)|\leq C(|\delta |+|b|)(|\tilde{\delta}-\delta _{0}|+|\tilde{b}%
-b_{0}|).
\end{equation*}

d) there is $C>0$ such that for $\psi _{i\delta x}=I(\tau _{i})-E[I(\tau
_{i})]-\delta ^{2}\{\tau _{i}-E[\tau _{i}]\}$ and all $J,$%
\begin{equation*}
(J+1)E[1(\tau _{i}<1/(J+1))\psi _{i\delta x}^{2}]\geq C.
\end{equation*}

e) Each of the following converge to zero: $\sqrt{n}J\varepsilon _{pn}^{2}$, 
$nJ^{3}/S,$ $J^{7/2}K/(\sqrt{S}\Delta ),$ $J^{7/2}K\Delta ,$ $%
J^{7/2}K^{3/2}\varepsilon _{pn},$ $J^{5/2}K^{-s_{\alpha }}$
\end{assumption}

Part a) is Frechet differentiability of the CDF of $\tau _{i}$ in the drift
and boundary, b) is implied by mean square continuity of the derivative and
the Riesz representation Theorem, and c) is continuity of the functional
derivative $D$ in $\delta $ and $b$. The test statistic will continue to be
asymptotically chi-squared for a stronger norm for $b$ under corresponding
stronger rate conditions for $J$, $K$, and $\Delta .$

\section{Proofs from Section \protect\ref{sec:metrics}}

We will use two Lemmas on the asymptotic behavior of quadratic forms to
prove the properties of the test statistic. For the first Lemma let $h_{i}$
be a $J\times 1$ vector of random variables with $E\left[ h_{i}\right] =0$
and $h_{1},\ldots ,h_{n}$ i.i.d. Let 
\begin{equation*}
\Omega =E\left[ h_{i}h_{i}^{\prime }\right] ,\;\bar{h}=\frac{1}{n}%
\sum_{i}h_{i}.
\end{equation*}%
Consider $\hat{h}$ that is approximately equal to $\bar{h}$ in the sense
that $\hat{h}-\bar{h}$ is small. Also consider an estimator $\hat{\Omega}$
of $\Omega $ and let $\left\Vert A\right\Vert =\sqrt{{tr}\left( A^{\prime
}A\right) }$ be the $L_{2}$ norm on matrices.

\begin{lemma}
\label{lem:initial} If i) $\lambda _{\min }\left( \Omega \right) \geq c>0$,
ii) $J^{-1/2}\sqrt{n}{tr}\left( \Omega \right) ^{1/2}\left\Vert \hat{h}-\bar{%
h}\right\Vert \overset{p}{\longrightarrow }0,$ iii) $J^{-1/2}{tr}\left(
\Omega \right) \left\Vert \hat{\Omega}-\Omega \right\Vert \overset{p}{%
\longrightarrow }0,$ and iv) $E\left[ \left( h_{i}^{\prime }h_{i}\right) ^{2}%
\right] /nJ\longrightarrow 0$ then for the $1-\alpha $ quantile $c\left(
\alpha ,J\right) $ of a chi-square distribution with $J$ degrees of freedom%
\begin{equation*}
\Pr \left( n\hat{h}^{\prime }\hat{\Omega}^{-1}\hat{h}\geq c\left( \alpha
,J\right) \right) \longrightarrow \alpha .
\end{equation*}
\end{lemma}

\noindent \textbf{Proof}: By i) we have $\lambda _{\min }\left( \Omega
\right) \geq c$, so that $J^{-1/2}{tr}\left( \Omega \right) ^{1/2}\geq c.$
Then iii) implies $\left\Vert \hat{\Omega}-\Omega \right\Vert \overset{p}{%
\longrightarrow }0$ and hence w.p.a.1,%
\begin{equation*}
\lambda _{\min }\left( \hat{\Omega}\right) \geq c.
\end{equation*}%
Since this event occurs w.p.a.1 we can assume it is true henceforth. Define 
\begin{equation*}
T_{1}=n^{\prime }\hat{h}\left( \hat{\Omega}^{-1}-\Omega ^{-1}\right) \hat{h}%
,\;T_{2}=n\left[ \hat{h}^{\prime }\Omega ^{-1}\hat{h}-\bar{h}^{\prime
}\Omega ^{-1}\bar{h}\right] 
\end{equation*}%
Note that $E[n\left\Vert \bar{h}\right\Vert ^{2}]=nE[\bar{h}^{\prime }\bar{h}%
]=tr(\Omega ).$Then by the Markov inequality we have 
\begin{equation*}
\sqrt{n}\left\Vert \bar{h}\right\Vert =O_{p}(tr(\Omega )^{1/2}).
\end{equation*}%
Also by ii) $\sqrt{n}\left\Vert \hat{h}-\bar{h}\right\Vert \leq
CJ^{-1/2}tr(\Omega )^{1/2}\sqrt{n}\left\Vert \hat{h}-\bar{h}\right\Vert 
\overset{p}{\longrightarrow }0.$ Then by the triangle inequality%
\begin{equation*}
\sqrt{n}\left\Vert \hat{h}\right\Vert \leq \sqrt{n}\left\Vert \bar{h}%
\right\Vert +\sqrt{n}\left\Vert \hat{h}-\bar{h}\right\Vert =O_{p}(tr(\Omega
)^{1/2}).
\end{equation*}%
It therefore follows that 
\begin{eqnarray*}
\left\vert T_{1}\right\vert  &=&\left\vert n\hat{h}^{\prime }\hat{\Omega}%
^{-1}\left( \Omega -\hat{\Omega}\right) \Omega ^{-1}\hat{h}\right\vert \leq
\left\Vert \sqrt{n}\hat{h}^{\prime }\hat{\Omega}^{-1}\right\Vert \left\Vert 
\hat{\Omega}-\Omega \right\Vert \left\Vert \sqrt{n}\hat{h}^{\prime }\Omega
^{-1}\right\Vert \leq cn\left\Vert \hat{h}\right\Vert ^{2}\left\Vert \hat{%
\Omega}-\Omega \right\Vert  \\
&=&O_{p}(tr(\Omega ))\left\Vert \hat{\Omega}-\Omega \right\Vert
=o_{p}(J^{1/2}).
\end{eqnarray*}%
Similarly we have 
\begin{eqnarray*}
\left\vert T_{2}\right\vert  &=&n\left\vert \left( \hat{h}-\bar{h}\right)
^{\prime }\Omega ^{-1}\hat{h}+\bar{h}^{\prime }\Omega ^{-1}\left( \hat{h}-%
\bar{h}\right) \right\vert \leq n(\left\Vert \hat{h}-\bar{h}\right\Vert
(\left\Vert \hat{h}\right\Vert +\left\Vert \bar{h}\right\Vert )) \\
&=&O_{p}(tr(\Omega )^{1/2}\sqrt{n}\left\Vert \hat{h}-\bar{h}\right\Vert
)=o_{p}(J^{1/2}).
\end{eqnarray*}%
It then follows by the triangle inequality that%
\begin{equation*}
n^{\prime }\hat{h}\hat{\Omega}^{-1}\hat{h}-n\bar{h}\Omega ^{-1}\bar{h}%
=T_{1}+T_{2}=o_{p}(J^{1/2}).
\end{equation*}%
In addition, by iv) and Lemma A.15 of \cite{NeweyWindmeijer09},%
\begin{equation*}
\frac{n\bar{h}^{\prime }\Omega ^{-1}\bar{h}-J}{\sqrt{2J}}\overset{d}{%
\longrightarrow }N\left( 0,1\right) .
\end{equation*}%
Also, by standard results for the chi-squared distribution, as $J\rightarrow
\infty $ we have $\left( c\left( \alpha ,J\right) -J\right) /\sqrt{2J}$
converges to the $1-\alpha $ quantile of a $N\left( 0,1\right) $. Hence%
\begin{equation*}
\Pr \left( n\bar{h}^{\prime }\Omega ^{-1}\bar{h}\geq c\left( \alpha
,J\right) \right) =\Pr \left( \frac{n\bar{h}^{\prime }\Omega ^{-1}\bar{h}-J}{%
\sqrt{2J}}\geq \frac{c\left( \alpha ,J\right) -J}{\sqrt{2J}}\right)
\longrightarrow \alpha .
\end{equation*}%
The conclusion then follows by the Slutzky Lemma. Q.E.D.

\bigskip

The next Lemma gives a rate of growth for the number of simulation draws to
ensure that the limiting distribution of the test statistic based on $\hat{m}%
_{S}$ is the same as that based on $\hat{m}=\int m\left( \tau _{s}\left( 
\hat{\delta},\hat{b}\right) \right) dF\left( s\right) .$

Let $h_{s}$ be simulated moments. \ Then we have

\begin{lemma}
If $\max\limits_{1\leq j\leq J}\sup\limits_{\tau >0}\left\vert m_{jJ}\left(
\tau \right) \right\vert \leq C\sqrt{J}$ and $nJ{tr}\left( \Omega \right)
/S\longrightarrow 0$ then 
\begin{equation*}
J^{-1/2}\sqrt{n}{tr}\left( \Omega \right) ^{1/2}\left\Vert \hat{m}_{S}-\hat{m%
}\right\Vert \overset{p}{\longrightarrow }0,
\end{equation*}
\end{lemma}

\noindent \textbf{Proof:} Let $Z=\left( \left( \gamma _{1},\tau _{1}\right)
,\ldots ,\left( \gamma _{n},\tau _{n}\right) \right) $ denote the data. Note
that by definition, $E[\hat{m}_{S}|Z]=\hat{m}$. Then for any constant $\ell $
\begin{equation*}
\lim {Prob}\left( \left\Vert \hat{m}_{S}-\hat{m}\right\Vert >\ell \right) =E%
\left[ \Pr \left( \left\Vert \hat{m}_{S}-\hat{m}\right\Vert >\ell \mid
Z\right) \right] .
\end{equation*}%
By the Markov inequality%
\begin{align*}
\Pr \left( \left\Vert \hat{m}_{S}-\hat{m}\right\Vert >\ell \mid Z\right) &
=\Pr \left( \left\Vert \hat{m}_{S}-\hat{m}\right\Vert ^{2}>\ell ^{2}\mid
Z\right) \leq E\left[ \sum_{j=1}^{J}\left( \hat{m}_{Sj}-\hat{m}_{j}\right)
^{2}\mid Z\right] /\ell ^{2} \\
& \leq \frac{1}{S}\sum_{j=1}^{J}E\left[ \hat{m}_{j}\left( \tau _{s}\left( 
\hat{\delta},\hat{\beta}\right) \right) ^{2}\mid Z\right] /\ell ^{2}\leq 
\frac{C^{2}J^{2}}{S\ell ^{2}}.
\end{align*}%
By iterated expectations we then have 
\begin{equation*}
\Pr \left( \left\Vert \hat{m}_{S}-\hat{m}\right\Vert >\ell \right) \leq 
\frac{C^{2}J^{2}}{S\ell ^{2}}.
\end{equation*}%
Let $\ell =J^{1/2}{tr}\left( \Omega \right) ^{-1/2}n^{-1/2}\varepsilon .$
Then 
\begin{align*}
& \Pr \left( J^{-1/2}{tr}\left( \Omega \right) ^{1/2}\sqrt{n}\left\Vert \hat{%
m}_{s}-\hat{m}\right\Vert \geq \varepsilon \right) =\Pr \left( \left\Vert 
\hat{m}_{s}-\hat{m}\right\Vert \geq \ell \right) \leq C^{2}J^{2}\left[ SJ{tr}%
\left( \Omega \right) ^{-1}n^{-1}\varepsilon ^{2}\right] ^{-1} \\
& =\frac{J^{2}{tr}\left( \Omega \right) n}{SJ\varepsilon ^{2}}=\frac{nJ{tr}%
\left( \Omega \right) }{S}\frac{1}{\varepsilon ^{2}}\longrightarrow 0.
\end{align*}%
Q.E.D.

\bigskip

We next give a uniform convergence rate for $\hat{p}(t)$. For notational
simplicity we let $p(t):=p^{xy}(t).$

\bigskip

\textbf{Lemma 4}: If Assumptions 2 and 3 are satisfied then 
\begin{equation*}
\sup_{t}\left\vert \hat{p}(t)-p(t)\right\vert =O_{p}(\sqrt{\frac{K\ln (K)}{n}%
}+K^{-s}).
\end{equation*}

\textbf{Proof:} Follows from Theorem 4.3 and Comments 4.5 and 4.6 of \cite%
{Bellonietal15}. $Q.E.D.$

\bigskip

We next give an asymptotic expansion for $\hat{\delta}$. Define%
\begin{eqnarray*}
I\left( p\right)  &=&p\ln \left( \frac{p}{1-p}\right) +\left( 1-p\right) \ln
\left( \frac{1-p}{p}\right) =\left( 1-2p\right) \ln \left( \frac{1-p}{p}%
\right) , \\
\psi _{i}^{\delta } &=&\frac{1}{2E[\tau _{i}]\delta }\left\{
I(p_{i})-I_{0}+I_{p}(p_{i})(\gamma _{i}-p_{i})-\delta ^{2}(\tau _{i}-E[\tau
_{i}])\right\} .
\end{eqnarray*}

\textbf{Lemma 5}: If Assumptions 2 and 3 are satisfied and $\sqrt{n}%
\varepsilon _{pn}^{2}\longrightarrow 0$ then%
\begin{equation*}
\hat{\delta}-\delta =\frac{1}{n}\sum_{i}\psi _{i}^{\delta
}+O_{p}(\varepsilon _{pn}^{2})=\frac{1}{n}\sum_{i}\psi _{i}^{\delta
}+o_{p}(1/\sqrt{n})=O_{p}(1/\sqrt{n}).
\end{equation*}%
\textbf{Proof:} Equation (4) and Assumption 3 imply that $p(t)$ is bounded
away from zero and one. It then follows from Lemma 4 that with probability
approaching one (w.p.a.1) there is $\varepsilon >0$ with $\varepsilon \leq $ 
$\hat{p}(t)\leq 1-\varepsilon .$ It is straightforward to check that $I(p)$
is twice continuously differentiable in $p\in (0,1)$ with first and second
deriatives that are bounded when $p$ is bounded away from zero and one. It
then follows by an expansion and Lemma 4 that%
\begin{equation*}
I\left( \hat{p}_{i}\right) =I\left( p_{i}\right) +I_{p}(p_{i})\left( \hat{p}%
_{i}-p_{i}\right) +\hat{R}_{i},\left\vert \hat{R}_{i}\right\vert \leq C|\hat{%
p}_{i}-p_{i}|^{2}.
\end{equation*}%
Therefore we have%
\begin{equation*}
\hat{I}=\frac{1}{n}\sum_{i}I\left( \hat{p}_{i}\right) =\frac{1}{n}%
\sum_{i}[I\left( p_{i}\right) +I_{p}\left( p_{i}\right) \left( \hat{p}%
_{i}-p_{i}\right) ]+\hat{R},\text{ }\hat{R}=O_{p}(\varepsilon _{pn}^{2}).
\end{equation*}%
Define 
\begin{eqnarray*}
\Gamma  &=&(\gamma _{1},...,\gamma _{n})^{\prime },\text{ }%
P=(p_{1},...,p_{n})^{\prime },\text{ }Q=[q^{K}(G_{1}),...,q^{K}(G_{n})]^{%
\prime },\text{ }I_{p}=(I_{p}(p_{1}),...,I_{p}(p_{n})), \\
H &=&I-Q(Q^{\prime }Q)^{-}Q.
\end{eqnarray*}%
Note that derivatives of $I_{p}(p)$ to any order are bounded on $%
[\varepsilon ,1-\varepsilon ],$ so that by the fact that the approximation
rate of a general $s$ differentiable function by a b-spline of at least
order $s-1$ is $K^{-s}$ we have%
\begin{equation*}
\frac{1}{n}P^{\prime }HP=O(K^{-2s}),\text{ }\frac{1}{n}I_{p}^{\prime
}HI_{p}=O(K^{-2s}).
\end{equation*}%
Note also that 
\begin{equation*}
\frac{1}{n}\sum_{i}I_{p}\left( p_{i}\right) \left( \hat{p}_{i}-p_{i}\right) -%
\frac{1}{n}\sum_{i}I_{p}\left( p_{i}\right) \left( \gamma _{i}-p_{i}\right)
=-\frac{1}{n}I_{p}^{\prime }H\Gamma 
\end{equation*}%
Furthermore, 
\begin{equation*}
E[-\frac{1}{n}I_{p}^{\prime }H\Gamma |\tau _{1},...,\tau _{n}]=-\frac{1}{n}%
I_{p}^{\prime }HP=O(K^{-2s}),\text{ }Var(-\frac{1}{n}I_{p}^{\prime }H\Gamma
|\tau _{1},...,\tau _{n})\leq \frac{1}{n^{2}}I_{p}^{\prime }HI_{p}=O(\frac{%
K^{-2s}}{n}).
\end{equation*}%
Then by $2K^{-s}/\sqrt{n}\leq 1/n+K^{-2s}\leq \varepsilon _{pn}^{2}$ it
follows that%
\begin{equation*}
\frac{1}{n}\sum_{i}I_{p}\left( p_{i}\right) \left( \hat{p}_{i}-p_{i}\right) -%
\frac{1}{n}\sum_{i}I_{p}\left( p_{i}\right) \left( \gamma _{i}-p_{i}\right)
=O_{p}(\frac{K^{-s}}{\sqrt{n}}+K^{-2s})=O_{p}(\varepsilon _{pn}^{2}).
\end{equation*}%
Then by the triangle inequality%
\begin{equation*}
\hat{I}=\frac{1}{n}\sum_{i}I\left( \hat{p}_{i}\right) =\frac{1}{n}%
\sum_{i}[I\left( p_{i}\right) +I_{p}\left( p_{i}\right) \left( \gamma
_{i}-p_{i}\right) ]+O_{p}(\varepsilon _{pn}^{2}).
\end{equation*}%
Note that for $\delta (I,\tau )=\sqrt{I/\tau },$%
\begin{equation*}
\frac{\partial \delta (I,\tau )}{\partial I}=\frac{1}{2\delta (I,\tau )\tau }%
,\text{ }\frac{\partial \delta (I,\tau )}{\partial \tau }=-\frac{\delta
(I,\tau )}{2\tau }.
\end{equation*}%
The conclusion then follows by the usual delta method argument. $Q.E.D.$

\bigskip

Next for any $\alpha (\tau )$ define%
\begin{equation*}
\psi _{i}^{\alpha }=-\delta ^{-1}\{E[\alpha (\tau _{i})b(\tau _{i})]\psi
_{i}^{\delta }+\frac{\alpha (\tau _{i})}{p(\tau _{i})[1-p(\tau _{i})]}%
(\gamma _{i}-p_{i})\}.
\end{equation*}

The next result gives a rate of convergence for the boundary estimator $\hat{%
b}(t)$ and a uniform expansion for a mean square continuous linear
functional of $\hat{b}(t)$ 

\bigskip 

\textbf{Lemma 6:} If there is a constant $C$ such that $\alpha (G^{-1}(g))$
is continuously differentiable of order $s$ with $\left\vert d\alpha
(G^{-1}(g))/dg\right\vert \leq C$ on $[0,1],$ then $\sup_{t}|\hat{b}%
(t)-b(t)|=O_{p}(\varepsilon _{pn})$ and 
\begin{equation*}
\int \alpha (\tau )\{\hat{b}(\tau )-b(\tau )\}F_{0}(d\tau )=\frac{1}{n}%
\sum_{i}\psi _{i}^{\alpha }+O_{p}(\varepsilon _{np}^{2}),
\end{equation*}%
uniformly in $\alpha $.

\bigskip

\textbf{Proof:} Note that for $b(\delta ,p)=\delta ^{-1}\ln (p/[1-p]),$ 
\begin{equation*}
\frac{\partial b(\delta ,p)}{\partial \delta }=\frac{-b(\delta ,p)}{\delta },%
\text{ }\frac{\partial b(\delta ,p)}{\partial p}=\frac{1}{\delta p(1-p)}.
\end{equation*}%
Then by Lemma 5, a delta method argument similar to that used in the proof
of Lemma 5, and $\hat{\delta}=\delta +O_{p}(1/\sqrt{n})$ we have%
\begin{equation*}
\hat{b}\left( t\right) =b(t)-b(t)\frac{[\hat{\delta}-\delta ]}{\delta }+%
\frac{1}{\delta p(t)[1-p(t)]}[\hat{p}(t)-p(t)]+\hat{R}(t),\text{ }%
\sup_{t}\left\vert \hat{R}(t)\right\vert =O_{p}(\varepsilon _{pn}^{2}).
\end{equation*}%
The first conclusion then follows by $b(t)$ bounded, which implies $p(t)$ is
bounded away from zero and one, and by Lemma 5. To show the second
conclusion note that for any bounded $a(t)$ it follows by the proof of
Corollary 10 of \cite{IchimuraNewey18} that%
\begin{equation*}
\int a(\tau )[\hat{p}(\tau )-p(\tau )]F_{0}(d\tau )=\frac{1}{n}%
\sum_{i}a(\tau _{i})[\gamma _{i}-p_{i}]+O_{p}(\varepsilon _{pn}^{2}),
\end{equation*}%
uniformly in $a(\tau )$ with uniformly bounded derivatives to order $s.$ Let 
$a(\tau )=\alpha (\tau )/\{\delta p(t)[1-p(t)]\}.$ By plugging in the above
expansion for $\hat{b}(t)$ and using boundedness of $\alpha (\tau )$ we
obtain%
\begin{eqnarray*}
&&\int \alpha (\tau )\{\hat{b}(\tau )-b(\tau )\}F_{0}(d\tau ) \\
&=&-\delta ^{-1}\{E[\alpha (\tau _{i})b(\tau _{i})](\hat{\delta}-\delta
)+\int a(\tau )[\hat{p}(\tau )-p(\tau )]F_{0}(d\tau )+\int \alpha (\tau )%
\hat{R}(\tau )F_{0}(d\tau ). \\
&=&\frac{1}{n}\sum_{i}\psi _{i}^{\alpha }+O_{p}(\varepsilon _{np}^{2})+\int
\alpha (\tau )\hat{R}(\tau )F_{0}(d\tau )=\frac{1}{n}\sum_{i}\psi
_{i}^{\alpha }+O_{p}(\varepsilon _{np}^{2}).\text{ }Q.E.D..
\end{eqnarray*}

\bigskip

Proof of Theorem 4: We first show that conditions i)-iv) of Lemma 2 are
satisfied. Let%
\begin{eqnarray*}
h_{ji} &=&m_{ji}-E[m_{ji}]+M_{\delta j}\psi _{i}^{\tau }+\alpha _{j0}(\tau
_{i})(\gamma _{i}-p_{i}),\text{ } \\
\psi _{i}^{\tau } &=&\frac{1}{2\delta E[\tau _{i}]}\{I(p_{i})-I_{0}-\delta
^{2}(\tau _{i}-E[\tau _{i}])\},\text{ } \\
M_{\delta j} &=&\sqrt{J}(D_{0\tau _{j+1}}^{\delta }-D_{0\tau _{j}}^{\delta
}-\delta ^{-1}E[\{\alpha _{0,\tau _{j+1}}(\tau _{i})-\alpha _{0,\tau
_{j}}(\tau _{i})\}b(\tau _{i})]) \\
\alpha _{j0}(\tau _{i}) &=&M_{\delta j}\frac{1}{2E[\tau _{i}]\delta }%
I_{p}(p_{i})+\frac{\sqrt{J}[\alpha _{0,\tau _{j+1}}(\tau _{i})-\alpha
_{0,\tau _{j}}(\tau _{i})]}{\delta p_{i}[1-p_{i}]}.
\end{eqnarray*}%
Also let 
\begin{eqnarray*}
h_{i} &=&(h_{i1},...,h_{iJ})^{\prime }=m_{i}-E[m_{i}]+M_{\delta }\psi
_{i}^{\tau }+\alpha _{0}(\tau _{i})(\gamma _{i}-p_{i}), \\
M_{\delta } &=&(M_{\delta 1},...,M_{\delta J})^{\prime },\text{ }\alpha
_{0}(\tau )=(\alpha _{10}(\tau ),...,\alpha _{J0}(\tau ))^{\prime }, \\
\Omega  &=&E[h_{i}h_{i}^{\prime }],\text{ }V_{1}=Var(m_{i}+M_{\delta }\psi
_{i}^{\tau }),\text{ }V_{2}=E[\alpha _{0}(\tau _{i})\alpha _{0}(\tau
_{i})^{\prime }Var(\gamma _{i}|\tau _{i})].
\end{eqnarray*}%
Note that $\Omega =V_{1}+V_{2}$ by $E[\gamma _{i}|\tau _{i}]=p(\tau _{i})$.

To show condition i) of Lemma 2 it suffices to show that $\lambda _{\min
}(V_{1})\geq C,$ which we now proceed to show. Let 
\begin{equation*}
\tilde{m}_{i}=(\sqrt{J+1}\psi _{i}^{\tau },m_{i}^{\prime })^{\prime }.
\end{equation*}%
It follows in a straightforward way from Assumption 5 d) that%
\begin{equation*}
\lambda _{\min }(E[\tilde{m}_{i}\tilde{m}_{i}^{\prime }])\geq C.
\end{equation*}%
Also, for $B=[M_{\delta },I]$ we have%
\begin{equation*}
V_{1}=BE[\tilde{m}_{i}\tilde{m}_{i}^{\prime }]B^{\prime }.
\end{equation*}%
Therefore for any conformable vector $\lambda $ with $\lambda ^{\prime
}\lambda =1,$%
\begin{equation*}
\lambda ^{\prime }V_{1}\lambda =\frac{\lambda ^{\prime }BE[\tilde{m}_{i}%
\tilde{m}_{i}^{\prime }]B^{\prime }\lambda }{\lambda ^{\prime }BB^{\prime
}\lambda }\lambda ^{\prime }BB^{\prime }\lambda \geq C\lambda ^{\prime
}BB^{\prime }\lambda \geq C\lambda _{\min }(BB^{\prime })\geq C\lambda
_{\min }(I)=C.
\end{equation*}

We next show that condition ii) of the Lemma 2 is satisfied. Recall that%
\begin{equation*}
m_{jJ}(t)=\sqrt{J}1(\tau _{j,J}\leq t<\tau _{j+1,J}),\text{ }(j=1,...,J).
\end{equation*}%
Then taking epectations over the simulation, 
\begin{eqnarray*}
E[m_{jS}(\delta ,b)] &=&\bar{m}_{j}(\delta ,b)=\int m_{jJ}(\tau _{s}(\delta
,b))F_{s}(ds) \\
&=&\sqrt{J}[F(\tau _{j+1,J}|\delta ,b)-F(\tau _{j,J}|\delta ,b)],\text{ }%
(j=1,...,J).
\end{eqnarray*}%
From Assumption 5 let 
\begin{equation*}
\hat{D}_{j}(\tilde{\delta},\tilde{b})=D(\tilde{\delta},\tilde{b};\hat{\delta}%
,\hat{b},\tau _{j}),\text{ }D_{j}(\tilde{\delta},\tilde{b})=D(\tilde{\delta},%
\tilde{b};\delta ,b,\tau _{j}).
\end{equation*}%
By Assumption 5 a) and Lemma 5, 
\begin{eqnarray*}
\bar{m}_{j}(\hat{\delta},\hat{b})-\bar{m}_{j}(\delta ,b) &=&\sqrt{J}[D_{j+1}(%
\hat{\delta}-\delta ,\hat{b}-b)-D_{j}(\hat{\delta}-\delta ,\hat{b}-b)]+\hat{R%
}_{j}, \\
\left\vert \hat{R}_{j}\right\vert  &\leq &\sqrt{J}2C[(\hat{\delta}-\delta
)^{2}+\sup_{t}|\hat{b}(t)-b(t)|^{2}]=O_{p}(\sqrt{J}\varepsilon _{pn}^{2}),
\end{eqnarray*}%
uniformly in $j.$ By Assumption 5 b) and Lemmas 5 and 6,%
\begin{eqnarray*}
&&\sqrt{J}[D_{j+1}(\hat{\delta}-\delta ,\hat{b}-b)-D_{j}(\hat{\delta}-\delta
,\hat{b}-b)] \\
&=&\sqrt{J}[(D_{0\tau _{j+1}}^{\delta }-D_{0\tau _{j}}^{\delta })(\hat{\delta%
}-\delta )+\int \{\alpha _{0,\tau _{j+1}}(\tau )-\alpha _{0,\tau _{j}}(\tau
)\}\{\hat{b}(\tau )-b(\tau )\}F_{0}(d\tau )] \\
&=&\sqrt{J}[(D_{0\tau _{j+1}}^{\delta }-D_{0\tau _{j}}^{\delta })\{\frac{1}{n%
}\sum_{i}\psi _{i}^{\delta }+O_{p}(\varepsilon _{pn}^{2})\}] \\
&&-\sqrt{J}\delta ^{-1}E[\{\alpha _{0,\tau _{j+1}}(\tau _{i})-\alpha
_{0,\tau _{j}}(\tau _{i})\}b(\tau _{i})])\left( \frac{1}{n}\sum_{i}\psi
_{i}^{\delta }\right)  \\
&&+\sqrt{J}\frac{1}{n}\sum_{i}\frac{[\alpha _{0,\tau _{j+1}}(\tau
_{i})-\alpha _{0,\tau _{j}}(\tau _{i})]}{\delta p_{i}[1-p_{i}]}(\gamma
_{i}-p_{i})+\sqrt{J}O_{p}(\varepsilon _{pn}^{2}) \\
&=&\frac{1}{n}\sum_{i}h_{ji}+O_{p}(\sqrt{J}\varepsilon _{pn}^{2})
\end{eqnarray*}%
Then by $tr(\Omega )^{1/2}=O(J)$ we have%
\begin{equation*}
J^{-1/2}\sqrt{n}{tr}\left( \Omega \right) ^{1/2}\left\Vert \hat{h}-\bar{h}%
\right\Vert \leq CJ^{1/2}\sqrt{n}\left\Vert \hat{h}-\bar{h}\right\Vert \leq C%
\sqrt{n}\sqrt{J}O_{p}(\sqrt{J}\varepsilon _{pn}^{2}).
\end{equation*}%
Hypothesis ii) of Lemma 2 then follows by $\sqrt{n}J\varepsilon
_{pn}^{2}\longrightarrow 0,$ and by Lemma 3 and $nJ^{3}/S\longrightarrow 0.$

Next we verify hypothesis iii) of Lemma 2. Note that%
\begin{equation*}
\hat{M}_{\delta j}=\frac{\hat{m}_{j}(\hat{\delta}+\Delta ,\hat{\beta})-\hat{m%
}_{j}(\hat{\delta}-\Delta ,\hat{\beta})}{2\Delta }
\end{equation*}%
Let $\bar{m}_{j}\left( \delta ,\beta \right) =\int m_{j}\left( \tau
_{s}\left( \delta ,\beta \right) \right) F\left( ds\right) $ and 
\begin{equation*}
\bar{M}_{\delta j}=\frac{\bar{m}_{j}\left( \hat{\delta}+\Delta ,\hat{\beta}%
\right) -\bar{m}_{j}\left( \hat{\delta}-\Delta ,\hat{\beta}\right) }{2\Delta 
}.
\end{equation*}%
By the simulations i.i.d. given $\hat{\delta},\hat{\beta}$ and $m_{jJ}(\tau
)\leq C\sqrt{J},$%
\begin{equation*}
E\left[ \left( \hat{M}_{\delta j}-\bar{M}_{\delta j}\right) ^{2}\mid \hat{%
\delta},\hat{\beta}\right] \leq \frac{CJ}{S\Delta ^{2}}.
\end{equation*}%
Then for $\bar{M}_{\delta }=(\bar{M}_{\delta 1},...,\bar{M}_{\delta
J})^{\prime }$ the Markov inequality gives%
\begin{equation*}
E\left[ \left\Vert \hat{M}_{\delta }-\bar{M}_{\delta }\right\Vert ^{2}\right]
\leq \frac{CJ^{2}}{S\Delta ^{2}},\text{ }\left\Vert \hat{M}_{\delta }-\bar{M}%
_{\delta }\right\Vert =O_{p}\left( \frac{J}{\sqrt{S}\Delta }\right) .
\end{equation*}%
Note that replacing $\hat{\delta}$ with $\hat{\delta}+\Delta $ in the
boundary estimator $\hat{b}$ gives $[\hat{\delta}/(\hat{\delta}+\Delta )]%
\hat{b}$ and replacing $\hat{\delta}$ with $\hat{\delta}-\Delta $ gives $[%
\hat{\delta}/(\hat{\delta}-\Delta )]\hat{b}.$ Also, 
\begin{equation*}
\frac{\hat{\delta}}{\hat{\delta}+\Delta }-1=\frac{-\Delta }{\hat{\delta}%
+\Delta },\text{ }\frac{\hat{\delta}}{\hat{\delta}-\Delta }-1=\frac{\Delta }{%
\hat{\delta}-\Delta }
\end{equation*}%
Let $\hat{D}_{j}(\delta ,b)=D(\delta ,b;\hat{\delta},\hat{b},j)$ and $%
D_{j}(\delta ,b)=D(\delta ,b;\delta _{0},b_{0},j)$ for true values $\delta
_{0}$ and $b_{0}.$ Then by Assumption 5 a),%
\begin{eqnarray*}
\bar{M}_{\delta j} &=&\frac{\bar{m}_{j}\left( \hat{\delta}+\Delta ,\hat{\beta%
}\right) -\bar{m}_{j}(\hat{\delta},\hat{\beta})-[\bar{m}_{j}\left( \hat{%
\delta}-\Delta ,\hat{\beta}\right) -\bar{m}_{j}(\hat{\delta},\hat{\beta})]}{%
2\Delta } \\
&=&\frac{\sqrt{J}[\hat{D}_{j+1}(\Delta ,\frac{-\Delta }{\hat{\delta}+\Delta }%
\hat{b})-\hat{D}_{j+1}(-\Delta ,\frac{\Delta }{\hat{\delta}-\Delta }\hat{b}%
;)]}{2\Delta }-\frac{\sqrt{J}[\hat{D}_{j}(\Delta ,\frac{-\Delta }{\hat{\delta%
}+\Delta }\hat{b})-\hat{D}_{j}(-\Delta ,\frac{\Delta }{\hat{\delta}-\Delta }%
\hat{b})]}{2\Delta }+\hat{R}_{j} \\
\left\vert \hat{R}_{j}\right\vert  &\leq &C\sqrt{J}\Delta ^{-1}(\Delta
^{2}+\left\vert \frac{\Delta }{\hat{\delta}+\Delta }\hat{b}\right\vert
^{2}+\left\vert \frac{\Delta }{\hat{\delta}-\Delta }\hat{b}\right\vert
^{2})\leq C\sqrt{J}\Delta (1+\left\vert \hat{b}\right\vert ^{2}).
\end{eqnarray*}%
We also have%
\begin{eqnarray*}
\sqrt{J}\frac{1}{\Delta }\hat{D}_{j+1}(\Delta ,\frac{-\Delta }{\hat{\delta}%
+\Delta }\hat{b}) &=&\sqrt{J}\hat{D}_{j+1}(1,\frac{-1}{\hat{\delta}+\Delta }%
\hat{b}), \\
\sqrt{J}|\hat{D}_{j+1}(1,\frac{-1}{\hat{\delta}+\Delta }\hat{b})-D_{j+1}(1,%
\frac{-1}{\hat{\delta}+\Delta }\hat{b})| &\leq &C\sqrt{J}\left\vert \frac{%
\hat{b}}{\hat{\delta}+\Delta }\right\vert (|\hat{\delta}-\delta |+|\hat{b}%
-b|)\leq C\sqrt{J}O_{p}(\varepsilon _{pn}).
\end{eqnarray*}%
Also,%
\begin{eqnarray*}
&&\sqrt{J}\left\vert D_{j+1}(1,\frac{-1}{\hat{\delta}+\Delta }\hat{b}%
)-D_{0\tau _{j+1}}^{\delta }+\frac{1}{\delta }\int \alpha _{0,\tau
_{j+1}}(\tau )b(\tau )F_{0}(d\tau )\right\vert  \\
&\leq &C\sqrt{J}(|\hat{\delta}-\delta |+|\hat{b}-b|)=\sqrt{J}%
O_{p}(\varepsilon _{pn}).
\end{eqnarray*}%
Applying an analogous set of inequalities to other terms and collecting
remainders gives%
\begin{equation*}
\left\vert \bar{M}_{\delta j}-M_{\delta j}\right\vert \leq C\sqrt{J}(\Delta
+O_{p}(\varepsilon _{pn})).
\end{equation*}%
Combining results and stacking over $j$ then give%
\begin{equation*}
\left\Vert \hat{M}_{\delta }-M_{\delta }\right\Vert =O_{p}(J(\frac{1}{\sqrt{S%
}\Delta }+\Delta +\varepsilon _{pn})).
\end{equation*}

Next, for $\hat{\psi}_{i}^{\tau }=\left( 2\hat{\delta}\bar{\tau}\right)
^{-1}[\hat{I}(\tau _{i})-\bar{I}-\hat{\delta}^{2}\{\tau _{i}-\bar{\tau}\}]$
it follows straighforwardly that%
\begin{equation*}
\frac{1}{n}\sum_{i=1}^{n}\left( \hat{\psi}_{i}^{\tau }-\psi _{i}^{\tau
}\right) ^{2}=O_{p}(\varepsilon _{pn}^{2}).
\end{equation*}%
Let $\tilde{V}_{1}=n^{-1}\sum_{i=1}^{n}\psi _{1i}\psi _{1i}^{\prime }$ and $%
\psi _{1i}=m_{i}-E[m_{i}]+M_{\delta }\psi _{i}^{\tau }.$ Note that%
\begin{eqnarray*}
\frac{1}{n}\sum_{i=1}^{n}\left\Vert \hat{\psi}_{1i}-\psi _{1i}\right\Vert
^{2} &\leq &\left\Vert \bar{m}-E[m_{i}]\right\Vert ^{2}+\left\Vert \hat{M}%
_{\delta }-M_{\delta }\right\Vert ^{2}\frac{1}{n}\sum_{i=1}^{n}\left\Vert 
\hat{\psi}_{i}^{\tau }\right\Vert ^{2}+\left\Vert M_{\delta }\right\Vert ^{2}%
\frac{1}{n}\sum_{i=1}^{n}(\hat{\psi}_{1i}-\psi _{1i})^{2} \\
&=&O_{p}(\frac{J^{2}}{n})+O_{p}(J^{2}(\frac{1}{\sqrt{S}\Delta }+\Delta
+\varepsilon _{pn})^{2})+O_{p}(J^{2}\varepsilon _{pn}^{2}) \\
&=&O_{p}(J^{2}(\frac{1}{\sqrt{S}\Delta }+\Delta +\varepsilon _{pn})^{2}).
\end{eqnarray*}%
Then by the Cauchy-Schwartz and triangle inequalities, 
\begin{eqnarray*}
\left\Vert \hat{V}_{1}-\tilde{V}_{1}\right\Vert  &\leq &\frac{1}{n}%
\sum_{i=1}^{n}\left\Vert \hat{\psi}_{1i}-\psi _{1i}\right\Vert ^{2}+\sqrt{%
\frac{1}{n}\sum_{i=1}^{n}\left\Vert \hat{\psi}_{1i}-\psi _{1i}\right\Vert
^{2}}\sqrt{\frac{1}{n}\sum_{i=1}^{n}\left\Vert \psi _{1i}\right\Vert ^{2}} \\
&=&O_{p}(J^{2}(\frac{1}{\sqrt{S}\Delta }+\Delta +\varepsilon _{pn})).
\end{eqnarray*}%
It follows similarly that $\left\Vert \tilde{V}_{1}-V_{1}\right\Vert
=O_{p}(J^{3/2}/\sqrt{n}),$ so by the triangle inequality,%
\begin{equation*}
\left\Vert \hat{V}_{1}-V_{1}\right\Vert =O_{p}(J^{2}(\frac{1}{\sqrt{S}\Delta 
}+\Delta +\varepsilon _{pn})).
\end{equation*}%
\bigskip 

Next we derive a convergence rate for $\left\Vert \hat{V}_{2}-V_{2}\right%
\Vert .$ Let%
\begin{eqnarray*}
D_{\beta } &=&E[\alpha _{0}(\tau _{i})q_{i}^{K}{}^{\prime }],\text{ }\Sigma
=E[q_{i}^{K}q_{i}^{K}{}^{\prime }],\text{ }\alpha _{K}(\tau _{i})=D_{\beta
}\Sigma ^{-1}q_{i}^{K}, \\
\Lambda  &=&E[q_{i}^{K}q_{i}^{K}{}^{\prime }(\gamma _{i}-p_{i})^{2}],\text{ }%
\bar{V}_{2}=D_{\beta }\Sigma ^{-1}\Lambda \Sigma ^{-1}D_{\beta }^{\prime
}=E[\alpha _{K}(\tau _{i})\alpha _{K}(\tau _{i})^{\prime }(\gamma
_{i}-p_{i})^{2}].
\end{eqnarray*}%
Note that by Assumption 5 b) and standard approximation properties of
splines 
\begin{equation*}
E[\{(\alpha _{0j}(\tau _{i})-\alpha _{Kj}(\tau _{i}))(\gamma
_{i}-p_{i})\}^{2}]\leq CE[\{\alpha _{0j}(\tau _{i})-\alpha _{Kj}(\tau
_{i})\}^{2}]\leq CK^{-2s_{\alpha }},
\end{equation*}%
for a constant $C$ that does not epend on $j.$ Then we have%
\begin{eqnarray*}
\left\Vert \bar{V}_{2}-V_{2}\right\Vert ^{2} &=&\sum_{j,\ell
=1}^{J}\{E[\alpha _{Kj}(\tau _{i})\alpha _{K\ell }(\tau _{i})(\gamma
_{i}-p_{i})^{2}]-E[\alpha _{0j}(\tau _{i})\alpha _{0\ell }(\tau _{i})(\gamma
_{i}-p_{i})^{2}]\}^{2} \\
&=&\sum_{j,\ell =1}^{J}\{E[\{\alpha _{Kj}(\tau _{i})-\alpha _{0j}(\tau
_{i})\}\alpha _{K\ell }(\tau _{i})(\gamma _{i}-p_{i})^{2}]+E[\alpha
_{0j}(\tau _{i})\{\alpha _{K\ell }(\tau _{i})-\alpha _{0\ell }(\tau
_{i})\}(\gamma _{i}-p_{i})^{2}]\}^{2} \\
&\leq &C\sum_{j,\ell =1}^{J}\{\sqrt{E[\{\alpha _{Kj}(\tau _{i})-\alpha
_{0j}(\tau _{i})\}^{2}]}\sqrt{E[\alpha _{K\ell }(\tau _{i})^{2}]} \\
&&+\sqrt{E[\{\alpha _{K\ell }(\tau _{i})-\alpha _{0\ell }(\tau _{i})\}^{2}]}%
\sqrt{E[\alpha _{0j}(\tau _{i})^{2}]}\}^{2} \\
&\leq &C\left( \sum_{j=1}^{J}E[\{\alpha _{Kj}(\tau _{i})-\alpha _{0j}(\tau
_{i})\}^{2}]\right) \left( \sum_{j=1}^{J}\{E[\alpha _{0j}(\tau
_{i})^{2}]+E[\alpha _{K\ell }(\tau _{i})^{2}]\}\right) \leq
CJ^{2}K^{-2s_{\alpha }}.
\end{eqnarray*}%
Taking square roots we have%
\begin{equation*}
\left\Vert \bar{V}_{2}-V_{2}\right\Vert \leq CJK^{-s_{\alpha }}.
\end{equation*}

Define%
\begin{equation*}
\bar{M}_{\beta jk}=\frac{\bar{m}_{j}\left( \hat{\delta},\hat{\beta}%
+e_{k}\Delta \right) -\bar{m}_{j}\left( \hat{\delta},\hat{\beta}-e_{k}\Delta
\right) }{2\Delta }.
\end{equation*}%
It follows similarly to $\left\Vert \hat{M}_{\delta }-\bar{M}_{\delta
}\right\Vert =\left\Vert \hat{M}_{\delta }-\bar{M}_{\delta }\right\Vert
=O_{p}\left( J/\sqrt{S}\Delta \right) $ that 
\begin{equation*}
\left\Vert \hat{M}_{\beta }-\bar{M}_{\beta }\right\Vert =O_{p}\left( J\sqrt{K%
}/\sqrt{S}\Delta \right) .
\end{equation*}

Next, let $\hat{p}_{\Delta k}(t)=\hat{p}(t)+\Delta q_{kK}(G(t))$ and $\hat{b}%
_{\Delta k}(t)=\hat{\delta}^{-1}\ln (\hat{p}_{\Delta k}(t)/[1-\hat{p}%
_{\Delta k}(t)]).$ By $\Delta \sqrt{K}\longrightarrow 0$ and $\sup_{G\in
\lbrack 0,1]}|q_{kK}(G)|\leq C\sqrt{K}$ it follows that $\sup_{t}\Delta
q_{kK}(G(t))\longrightarrow 0.$ Then w.p.a.1 we have%
\begin{equation*}
\hat{b}_{\Delta k}(t)=\hat{b}(t)+\frac{\Delta q_{kK}(G(t))}{\hat{\delta}\hat{%
p}(t)[1-\hat{p}(t)]}+\hat{R}_{k}(t,\Delta ),\text{ }\left\vert \hat{R}%
_{k}(t,\Delta )\right\vert \leq C\Delta ^{2}K.
\end{equation*}

Then we have%
\begin{eqnarray*}
\bar{M}_{\beta jk} &=&\frac{\bar{m}_{j}\left( \hat{\delta},\hat{\beta}%
+e_{k}\Delta \right) -\bar{m}_{j}(\hat{\delta},\hat{\beta})-[\bar{m}%
_{j}\left( \hat{\delta},\hat{\beta}-e_{k}\Delta \right) -\bar{m}_{j}(\hat{%
\delta},\hat{\beta})]}{2\Delta } \\
&=&\frac{\sqrt{J}[\hat{D}_{j+1}(0,\hat{b}_{\Delta k}-\hat{b})-\hat{D}%
_{j+1}(0,\hat{b}_{-\Delta k}-\hat{b})]}{2\Delta } \\
&&-\frac{\sqrt{J}[\hat{D}_{j}(0,\hat{b}_{\Delta k}-\hat{b})-\hat{D}_{j}(0,%
\hat{b}_{-\Delta k}-\hat{b})]}{2\Delta }+\hat{R}_{jk} \\
\left\vert \hat{R}_{jk}\right\vert  &\leq &C\sqrt{J}\Delta ^{-1}(\left\vert 
\hat{b}_{\Delta k}-\hat{b}\right\vert ^{2}+\left\vert \hat{b}_{-\Delta ,k}-%
\hat{b}\right\vert ^{2})\leq C\sqrt{J}\Delta K.
\end{eqnarray*}

We also have%
\begin{eqnarray*}
\sqrt{J}\frac{1}{\Delta }\hat{D}_{j+1}(0,\hat{b}_{\Delta k}-\hat{b}) &=&%
\sqrt{J}\hat{D}_{j+1}(0,\frac{\hat{b}_{\Delta k}-\hat{b}}{\Delta }), \\
\sqrt{J}|\hat{D}_{j+1}(0,\frac{\hat{b}_{\Delta k}-\hat{b}}{\Delta }%
)-D_{j+1}(0,\frac{\hat{b}_{\Delta k}-\hat{b}}{\Delta })| &\leq &C\sqrt{J}%
\left\vert \frac{\hat{b}_{\Delta k}-\hat{b}}{\Delta }\right\vert (|\hat{%
\delta}-\delta |+|\hat{b}-b|)\leq C\sqrt{J}\sqrt{K}O_{p}(\varepsilon _{pn}).
\end{eqnarray*}

In addition%
\begin{eqnarray*}
\sqrt{J}D_{j+1}(0,\frac{\hat{b}_{\Delta k}-\hat{b}}{\Delta };\delta ,b,\tau
_{j+1}) &=&\sqrt{J}D(0,\frac{q_{kK}(G(\cdot ))}{\hat{\delta}\hat{p}(\cdot
)[1-\hat{p}(\cdot )]};\delta ,b,\tau _{j+1})+\sqrt{J}\Delta D(0,\hat{R}%
_{k}(\cdot ,\Delta );\delta ,b,\tau _{j+1}) \\
&=&\sqrt{J}D(0,\frac{q_{kK}(G(\cdot ))}{\delta p(\cdot )[1-p(\cdot )]}%
;\delta ,b,\tau _{j+1})+\hat{R}_{jk}, \\
\left\vert \hat{R}_{jk}\right\vert  &\leq &\sqrt{J}\sqrt{K}O_{p}(\varepsilon
_{pn})+\sqrt{J}K\Delta .
\end{eqnarray*}%
Combining terms we have%
\begin{equation*}
\left\Vert \hat{M}_{\beta }-M_{\beta }\right\Vert =O_{p}(J\sqrt{K}/\sqrt{S}%
\Delta +JK\varepsilon _{pn}+JK^{3/2}\Delta )
\end{equation*}

Next, we have 
\begin{eqnarray*}
&&\left\Vert \hat{M}_{\delta }\frac{1}{2\hat{\delta}\bar{\tau}n}%
\sum_{i=1}^{n}I_{p}(\hat{p}_{i})q_{i}^{K}{}^{\prime }-M_{\delta }\frac{1}{%
2\delta E[\tau _{i}]}E[I_{p}(p_{i})q_{i}^{K\prime }]\right\Vert  \\
&\leq &\left\Vert \hat{M}_{\delta }-M_{\delta }\right\Vert \frac{1}{2\hat{%
\delta}\bar{\tau}}\left( \frac{1}{n}\sum_{i=1}^{n}I_{p}(\hat{p}%
_{i})^{2}\right) ^{1/2}\left( \frac{1}{n}\sum_{i=1}^{n}q_{i}^{K}{}^{\prime
}q_{i}^{K}\right) ^{1/2} \\
&&+\left\Vert M_{\delta }\right\Vert \left\Vert \frac{1}{2\hat{\delta}\bar{%
\tau}n}\sum_{i=1}^{n}I_{p}(\hat{p}_{i})q_{i}^{K}{}^{\prime }-\frac{1}{%
2\delta E[\tau _{i}]}E[I_{p}(p_{i})q_{i}^{K\prime }]\right\Vert  \\
&=&O_{p}(J\sqrt{K}(\frac{1}{\sqrt{S}\Delta }+\Delta +\varepsilon
_{pn}))+O_{p}(JK\varepsilon _{pn})=O_{p}(J\sqrt{K}(\frac{1}{\sqrt{S}\Delta }%
+\Delta +\sqrt{K}\varepsilon _{pn})).
\end{eqnarray*}%
Combining terms we then have

\begin{equation*}
\left\Vert \hat{D}_{\beta }-D_{\beta }\right\Vert =O_{p}(J\sqrt{K}/\sqrt{S}%
\Delta +JK\varepsilon _{pn}+JK^{3/2}\Delta ).
\end{equation*}

Next, for $\hat{\pi}=\hat{\Sigma}^{-1}\hat{D}_{\beta }$ and $\pi =\Sigma
^{-1}D_{\beta }$ note that $\hat{V}_{2}=\hat{\pi}^{\prime }\hat{\Lambda}\hat{%
\pi}$ and $\bar{V}_{2}=\pi ^{\prime }\Lambda \pi .$ Also we have%
\begin{equation*}
\hat{V}_{2}-\bar{V}_{2}=(\hat{\pi}-\pi )^{\prime }\hat{\Lambda}(\hat{\pi}%
-\pi )+2\pi ^{\prime }\hat{\Lambda}(\hat{\pi}-\pi )+\pi ^{\prime }(\hat{%
\Lambda}-\Lambda )\pi .
\end{equation*}%
By the law of large number for symmetric matrices, $\left\Vert \hat{\Sigma}%
-\Sigma \right\Vert _{op}=O_{p}\left( \sqrt{n^{-1}K\ln K}\right) =o_{p}(1),$
where $\left\Vert \cdot \right\Vert _{op}$ denotes the operator norm on
symmetric matrices. Then by the eigenvalues of $\Sigma $ bounded and bounded
away from zero, $\lambda _{\max }(\hat{\Sigma})=O_{p}(1)$ and $1/\lambda
_{\min }(\hat{\Sigma})=O_{p}(1).$ Let $\tilde{\Lambda}=\frac{1}{n}%
\sum_{i}q_{i}^{K}q_{i}^{K^{\prime }}\left( \gamma _{i}-p_{i}\right) ^{2}$.
Note that%
\begin{eqnarray*}
\hat{\Lambda}-\tilde{\Lambda} &=&\frac{1}{n}\sum_{i}q_{i}^{K}q_{i}^{K^{%
\prime }}\left[ \left( \gamma _{i}-\hat{p}_{i}\right) ^{2}-\left( \gamma
_{i}-p_{i}\right) ^{2}\right] \leq \frac{1}{n}\sum_{i}q_{i}^{K}q_{i}^{K^{%
\prime }}\left\vert \left( \gamma _{i}-\hat{p}_{i}\right) ^{2}-\left( \gamma
_{i}-p_{i}\right) ^{2}\right\vert  \\
&\leq &C\hat{\Sigma}\max_{i}\left\vert \hat{p}_{i}-p_{i}\right\vert =\hat{%
\Sigma}O_{p}\left( \varepsilon _{pn}\right) ,\text{ }\hat{\Lambda}-\tilde{%
\Lambda}\geq -C\hat{\Sigma}O_{p}(\varepsilon _{pn}).
\end{eqnarray*}%
Also by the law of large numbers for symmetric matrices $\left\Vert \tilde{%
\Lambda}-\Lambda \right\Vert _{op}=O_{p}\left( \sqrt{n^{-1}K\ln K}\right) .$
Therefore by the triangle inequality,%
\begin{equation*}
\left\Vert \hat{\Lambda}-\Lambda \right\Vert _{op}=O_{p}\left( \varepsilon
_{pn}\right) .
\end{equation*}%
It follows that $\lambda _{\max }(\hat{\Lambda})=O_{p}(1),$ $1/\lambda
_{\min }(\hat{\Lambda})=O_{p}(1),$ and for $\hat{\Upsilon}=\hat{\Lambda}%
-\Lambda ,$%
\begin{equation*}
\left\Vert \hat{\Upsilon}\right\Vert =\sqrt{tr(\hat{\Upsilon}^{2})}\leq C%
\sqrt{J}\left\Vert \hat{\Lambda}-\Lambda \right\Vert _{op}=O_{p}(\sqrt{J}%
\varepsilon _{pn}).
\end{equation*}%
Similarly we have $\left\Vert \hat{\Sigma}-\Sigma \right\Vert =O_{p}(K\sqrt{%
\ln (K)/n}).$ We also have $\left\Vert D_{\beta }\right\Vert \leq CJ\sqrt{K}.
$Then it follows that for $\varepsilon _{Dn}=J\sqrt{K}/\sqrt{S}\Delta
+JK\varepsilon _{pn}+JK^{3/2}\Delta $%
\begin{equation*}
\left\Vert \hat{\pi}-\pi \right\Vert \leq \left\Vert (\hat{D}_{\beta
}-D_{\beta })^{\prime }\hat{\Sigma}^{-1}\right\Vert +\left\Vert D_{\beta
}{}^{\prime }\hat{\Sigma}^{-1}(\Sigma -\hat{\Sigma})\Sigma ^{-1}\right\Vert
\leq O_{p}(\varepsilon _{Dn})+O_{p}(JK\sqrt{\ln (K)/n})=O_{p}(\varepsilon
_{Dn}).
\end{equation*}%
It then follows by the triangle inequality that 
\begin{eqnarray*}
\left\Vert \hat{V}_{2}-\bar{V}_{2}\right\Vert  &\leq &O_{p}(1)(\left\Vert 
\hat{\pi}-\pi \right\Vert ^{2}+\left\Vert \pi \right\Vert \left\Vert \hat{\pi%
}-\pi \right\Vert +\left\Vert \pi \right\Vert ^{2}\left\Vert \hat{\Lambda}%
-\Lambda \right\Vert ) \\
&=&O_{p}(J\sqrt{K}\varepsilon _{Dn}+J^{2}K^{2}\sqrt{\ln (K)/n})=O_{p}(J^{2}K/%
\sqrt{S}\Delta +J^{2}K^{3/2}\varepsilon _{pn}+J^{2}K\Delta ).
\end{eqnarray*}%
By the triangle inequality we then have%
\begin{equation*}
\left\Vert \hat{\Omega}-\Omega \right\Vert =O_{p}(J^{2}K/\sqrt{S}\Delta
+J^{2}K\Delta +J^{2}K^{3/2}\varepsilon _{pn}+JK^{-s_{\alpha }})
\end{equation*}%
It then follows that Assumption iii) is satsified by Assumption 5 e).

Finally, for Assumption iv) of Lemma A2, note that%
\begin{equation*}
\left( h_{i}^{\prime }h_{i}\right) ^{2}=\left(
\sum_{j=1}^{J}h_{ij}^{2}\right)
^{2}=\sum_{j=1}^{J}\sum_{k=1}^{K}h_{ij}^{2}h_{ik}^{2}\leq
CJ\sum_{j=1}^{J}h_{ij}^{4}\leq CJ^{4},
\end{equation*}%
so that%
\begin{equation*}
E\left[ \left( h_{i}^{\prime }h_{i}\right) ^{2}\right] /nJ\leq
CJ^{3}/n\longrightarrow 0.
\end{equation*}%
Therefore condition iv) is satisfied. Q.E.D.

\ifx\undefined\BySame
\newcommand{\BySame}{\leavevmode\rule[.5ex]{3em}{.5pt}\ }
\fi
\ifx\undefined\textsc
\newcommand{\textsc}[1]{{\sc #1}}
\newcommand{\emph}[1]{{\em #1\/}}
\let\tmpsmall\small
\renewcommand{\small}{\tmpsmall\sc}
\fi

\end{document}